\documentstyle[aps,epsfig,tighten,preprint]{revtex}
\begin{document}
\draft
\title{Lyapunov Exponent and the Solid-Fluid Phase Transition}
\author{Kyung-Hoon Kwon}
\address{Korea Basic Science Institute, Taejon 305-333, Korea}
\author{Byung-Yoon Park}
\address{Department of Physics, Chungnam National University, \\
Taejon 305-764, Korea}

%----------------------------- Title page --------------------------------
%\preprint{{\sl \today} \hskip 10.7cm Ver. 1.1} 
%\preprint{\parbox[b]{1.5in}{SNUTP-95/XXX \\ {\sl April 1995}} }

%\date{}
\maketitle
\begin{abstract}
We study changes in the chaotic properties of a many-body system
undergoing a solid-fluid phase transition. 
To do this, we compute the temperature 
dependence of the largest Lyapunov exponents 
$\lambda_{\mbox{\scriptsize max}}$ for both two- and three-dimensional
periodic systems of $N$-particles for various densities.  
The particles interact through a
soft-core potential.  The two-dimensional system exhibits an apparent
second-order phase transition as
indicated by a $\lambda$-shaped peak in the specific heat. 
The first derivative of $\lambda_{\mbox{\scriptsize max}}$ 
with respect to the temperature shows a peak at the same temperature. 
The three-dimensional system shows jumps, in both
system energy and $\lambda_{\mbox{\scriptsize max}}$,
at the same temperature, suggesting a first-order phase transition.
Relaxation phenomena in the phase-transition region are
analyzed by using the local time averages.
\end{abstract}

\newpage
%\narrowtext 
%%%%%%%%%%%%%%%%%%%%%%%%%%%%%%%%%%%%%%%%%%%%%%%%%%%%%%%%%%%%%%%%%
%%%%%%%%%%%%%%%%%%%%%%% Introduction  %%%%%%%%%%%%%%%%%%%%%%%%%%%
%%%%%%%%%%%%%%%%%%%%%%%%%%%%%%%%%%%%%%%%%%%%%%%%%%%%%%%%%%%%%%%%%

\section{Introduction}

Molecular dynamics (MD)\cite{MD} is a computer simulation
methods which relates the macroscopic properties of matter to a
microscopic description of the constituent particles' motion. In MD simulations
Newton's equations of motion are solved. Then, the thermodynmic quantities
of the system, such as
pressure and temperature, are obtained as time averages of
corresponding physical quantities. These time averages are fully
equivalent to statistical ensemble averages obtained by using the Monte Carlo (MC)
simulations. The dynamic method has been successfully applied to simulating not only
the static equilibrium systems, but also those in nonequilibrium.

Since the pioneering work of Hoover {\it et al.}\cite{HPB},
many investigations of the chaotic properties of the many-
particle systems have been carried out. 
Our own goal is to understand the relation between
the irreversible macroscopic behavior of the atomic systems and the
underlying microscopic theory with time-reversal symmetry. The MD
simulation provides a plausible clue to solving this long-standing problem.
The main
point is that the motion of the particles in a many-particle system is
Lyapunov unstable; that is, the phase trajectories starting from neighboring
initial points separate from one another exponentially in time, while they
explore only a restricted portion of phase space, forming a stable strange
attractor.

The spectrum of Lyapunov exponents $\{ \lambda_1, \cdots, \lambda_M\}$ 
is a powerful tool for the analysis of the properties of 
chaotic systems\cite{LE}. 
Lyapunov exponents measure the averaged exponential rates of
divergence or convergence of neighboring trajectories in phase space: 
the sum of the first $n$ Lyapunov exponents is defined 
by the exponential growth or shrinking rate of an $n$-dimensional 
phase-space volume. 
The resulting $\lambda_i(i=1,2,\cdots,M)$ are conventionally arranged 
in decreasing order. 
For chaotic systems the largest exponent, 
$\lambda_1$ (hereafter we will also denote
it as $\lambda_{\mbox{\scriptsize max}}$) is positive, so that neighboring
trajectories diverge exponentially. 
The Lyapunov dimension, following the conjecture of
Kaplan and York\cite{KY}, is a lower bound on the fractal dimension of the
above mentioned strange attractor. 
Furthermore, the sum of all positive Lyapunov exponents 
defines the Kolmogorov entropy\cite{BGS,SN}.

The most peculiar property of the many-particle systems is 
that they can be in different phases. 
A system can undergo a phase transition from one to another phase
when temperature or pressure is changed appropriately. 
Thus, it is interesting to see how the chaotic properties of a
system change during these phase transitions. 
Qualitative differences in the shapes of the Lyapunov spectra 
for fluids and solids were described in Ref.\cite{PH}. 
Further studies, described in Ref.\cite{BBP}, covered a wide range of
densities and temperatures, and led to the conclusion that the spectral
shape does not uniquely determine the phase of the system. 
A more quantitative analysis was carried out in Ref.\cite{DP}, 
where Lyapunov exponents for the correlated cell model\cite{AHW}, 
the Lorenz gas model\cite{MZ} and the Lennard-Jones fluid
were evaluated as a function of the density for various energies. 
In all the cases studied the $\lambda _{\mbox{\scriptsize max}}$ 
exhibited a maximum near the phase transition.

In this work we have the same interest. 
We study $D$-dimensional ($D$=2,3) equilibrium dense fluid models, 
where $N$-particles in a periodic box interacting with
 a soft-core repulsive potential. 
We evaluate $\lambda _{\mbox{\scriptsize max}}$ for these systems,
as a function of
the temperature, for various densities. 
In the two-dimensional case, 
a careful analysis of the specific heat, shows a 
$\lambda$-shaped peak, suggesting that the system undergoes 
a solid-fluid phase transition 
of the second order. 
Next, it is shown that the first derivative of 
$\lambda_{\mbox{\scriptsize max}}$ with
respect to the temperature has a peak at the same temperature. 
In the three-dimensional case $D$=3, both the system energy and 
$\lambda_{\mbox{\scriptsize max}}$ show jumps at the same temperature, 
implying that the phase transition is first-order. 
Such a phase transition takes place over a somewhat wide
range of temperature, 
where two phases coexist, so that the time
avergages require a much longer time to converge. 
With the help of the local time averages we can analyze 
relaxation phenomena in the phase-transition region.

In the following section II, 
we briefly describe the dense-fluid model and the
method for evaluating the Lyapunov exponents. 
The numerical results for a two-dimenional system with $N$=30 
and a three-dimensional system with $N$=32 are then presented and 
analysed in Sec. III and IV.  A conclusion follows.

%%%%%%%%%%%%%%%%%%%%%%%%%%%%%%%%%%%%%%%%%%%%%%%%%%%%%%%%%%%%%%%%%%%%%%%%%%%%%%
%%%%%%%%%%%%%%%%%%%%% MODEL for DENSE FLUIDS and SOLIDS %%%%%%%%%%%%%%%%%%%%%%
%%%%%%%%%%%%%%%%%%%%%%%%%%%%%%%%%%%%%%%%%%%%%%%%%%%%%%%%%%%%%%%%%%%%%%%%%%%%%%

\section{Model of Dense Fluids and Solids at Equilibrium}

The microscopic dynamics of dense fluids and solids at equilibrium can be
modeled by a Newtonian many-body system of $N$-particles in a periodic box.
The particles interact with each other through a short-range repulsive
pair potential. In this work, to optimize the numerical processes,
we adopt the short-range pair potential introduced in Ref.\cite{HPB}, 
\begin{equation}
\phi = 100 (1-r^2)^4, \hskip 5mm r < 1,
\end{equation}
truncated at the cutoff radius, $r=1$, where the first three derivatives
vanish. 
The smooth truncation at short range minimizes the errors 
associated with numerical integration\cite{HPB}. 
The repulsive part of this potential resembles that of the rare-gas system.
The fact that the potential is finite at the origin does not
cause any problem within the range of the system energy and density  
considered here. 
One may elaborate the model by employing a more realistic 
potential at the expense of additional notational complexity and reduced
computing speed.

By choosing the side length $L$ of the periodic box larger than twice the
interaction range, the problem becomes relatively simple. Among all the possible
pairs of particle $i$ and particle $j$, with its imaginary periodic
particles $j^\prime$, only the one with the shortest separation distance can
be included in evaluating the potential energy.

Let ${\bbox \Gamma}(t)$ be the phase space vector of $2DN$ variables describing
the motion of the particles 
\begin{equation}
\bbox{\Gamma}(t) \equiv (x_1, x_2, \cdots, x_N, y_1,\cdots,z_N; p_{x_1}, \cdots,
p_{z_N}).
\end{equation}
One may reduce the dimension of the phase-space vector by using the fact
that the center-of-mass motion of the system is trivial and further by using
the energy conservation. However, implementing this idea involves
 a lot of unnecessary
complication.

The Hamiltonian for the particle system is expressed as 
\begin{equation}
H = \sum_{i=1}^N \textstyle\frac12 p_i^2
 + \displaystyle \sum_{i>j} \phi(r_{ij})
\end{equation}
where $r_{ij}$ is the shortest distance between particle $i$ 
and particle $j$, $j^\prime$, $\cdots$ as explained above. 
Note that the summation for the
potential energy is not restricted to the pair potential 
between particles within the system box, 
including image particles. 
The Hamiltonian equations of motion 
\begin{equation}
\dot{q} = \frac{\partial H}{\partial p} , 
\hskip 2em 
\dot{p} = - \frac{\partial H}{\partial q}
\end{equation}
lead to a set of the first-order 
differential equations expressed in a form of 
\begin{equation}
\dot{\bbox \Gamma} = {\bbox G}({\bbox \Gamma}).  \label{r}
\label{eqm}\end{equation}
These equations describe the microscopic dynamics of the system in phase-space.
The macroscopic properties of the system can be studied through the
thermodynamical quantities defined by time averages. For example, the
temperature of the system is defined as 
\begin{equation}
\frac{D}{2} N T \equiv 
 \langle \sum_{i=1}^N \textstyle\frac12 p^2_i \rangle_t^{}
 = \displaystyle \lim_{\tau \rightarrow \infty} \int^\tau_0
  \!\!\! dt \sum_{i=1}^N \textstyle\frac12 p_i^2.
\end{equation}
Here, we use reduced units for 
which the mass of the particle, 
the interaction range of the potential, 
and Boltzmann's constant are unity.

Lyapunov exponents $(\lambda_1$, $\lambda_2$, $\cdots)$ measure the
long-term averaged exponential rates of divergence or convergence of
neighboring trajectories in phase space. They are arranged in decreasing
order, the first (largest) Lyapunov exponent, $\lambda_1$ (equivalently
denoted as $\lambda_{\mbox{\scriptsize max}}$ throughout this paper),
describes the exponential growth rate of the distance ($\ell_1$) between the
reference trajectory and the satellite trajectory 1, the sum of the first
two, $\lambda_1+\lambda_2$, describes that of the area $(a_{12})$ spanned by
the reference trajectory and the two satellite trajectories 1 and 2, and so
on. In this paper, we will consider only the largest exponent 
$\lambda_{\mbox{\scriptsize max}}$. It can be calculated by monitoring 
the length of a differential offset vector $\delta\bbox{\Gamma}$ 
in the tanget space to the reference trajectory. It presents a satelite 
trajectory infinitesimally separated from the reference one, with 
equations of motion derived from Eq.(\ref{eqm}) 
in a linearized form: 
\begin{equation}
\delta\dot{\bbox{\Gamma}}
 = \frac{\partial \bbox{G}}{\partial\bbox{\Gamma}} \cdot 
   \delta\bbox{\Gamma}
\end{equation}

%%%%%%%%%%%%%%%%%%%%%%%%%%%%%%%%%%%%%%%%%%%%%%%%%%%%%%%%%%%%%%%%%%%%
%%%%%%%%%%%%%%%%%%%%% Two-Dimensional System %%%%%%%%%%%%%%%%%%%%%%%
%%%%%%%%%%%%%%%%%%%%%%%%%%%%%%%%%%%%%%%%%%%%%%%%%%%%%%%%%%%%%%%%%%%%
\section{Two-Dimensional System}

We consider here a system of $N$=30 interacting particles moving
in a rectangular periodic box. 
The size and the shape of the periodic box is chosen
to contain $15(=5\times 3)$ primitive cells of the triangular lattice when
the particles are arranged into a configuration of the lowest potential
energy. (See Fig. 1a.) 
Finite-size effects 
are minimized by using the periodic boundary condition.  
Since the dimension of the primitive cell containing two particles is 
$d\times \sqrt{3}d$ with the lattice constant $d$, 
the particle number density $\rho$ is determined as $\rho = 2/(\sqrt{3}d^2)$
and the shape of the periodic box is close to a square
({\it i.e.\/}, $L_x:L_y=5:3\sqrt3$).
At $t$=0, 30 particles begin to move from a triangular lattice 
configuration, which is denoted by the solid circles in Fig. 1.
For $\rho > 2/\sqrt3 \sim 1.15$, 
six nearest neighboring particles can be found 
within the interaction range. 
It leads to a nonvanishing 
potential energy of $V_0=\phi (d)\times 6\times 30\times \frac 12$. 
In the following discussions, for a convenience
we will subtract $V_0$ from the total energy of the system. 
The velocities of the particles are chosen randomly with a fixed total kinetic
energy which determines the total energy $E$. 
Then, the equations of motion are solved numerically, using fourth-order 
Runge-Kutta method with the time step 0.001. 
This time step 0.001 is sufficiently small for all 
temperatures and densities studied 
in this work: the 
total energy is conserved with an accuracy of at least eight decimal digits.

In Fig. 1, we show two characteristic motions of a single particle in a
two-dimensional system of density $\rho =1.2$ at different system energies
$E/N=0.01$ and $E/N=1.5$.
We present the trajectories of a single particle for the first 100 
time units after the particles begin to move.
In case of $E/N=0.01$ (Fig.1a), 
each particle moves in a restricted area around its initial position.
Such motion defines a solid phase, where the system maintains  
an ordered configuration.  
On the other hand, in the system with $E/N=1.5$(Fig. 1b) 
a particle can wander over the all the space with no restriction,
defining the fluid phase.
The system may exhibit other phases such as gas or glass.
In any cases, there should be certain temperature(s) 
the system transfers from one phase to another.

Our interest here is relating the phase transition to  
the chaoticity of a system changing due to such a phase transition.
In Fig. 2, we present numerical results for the time-averaged 
kinetic energy per particle ($\langle K \rangle /N$ = temperature $T$) 
\it versus \rm the total energy per particle and the largest Lyapunov exponent 
\it versus \rm the temperature for the systems at two different densities. 
For each density, we carry out 80 simulations at every $E/N=0.05$ 
up to $E/N=4.0$.  
Open circles denote time averages taken over 
2000 time units ($\tau=2000$), excluding data up to 
first 100 time units. 
To corroborate convergence, we present time averages 
evaluated for $\tau=1000$ by the filled circles.
As for the time-averaged kinetic energies, the convergence is 
quite good. 
The filled circles can be hardly seen behind the open circles 
due to their small differences. 
We interpret the difference between them as a relative 
error of the numerical calculation. 
The temperatures are obtained with 0.1\% accuracy.
The Lyapunov exponents converge rather slowly.
Compared with the averages of $\tau=1000$, those of $\tau=2000$ 
reveal relative errors up to 1\%.

The data do not seem to exhibit any sudden changes 
indicating a phase transition.
In case of $\rho=1.0$, the system energy per particle 
$E/N$ is almost linear in temperature $T$ and 
$\lambda_{\mbox{\scriptsize max}}$ can be fitted by a single 
curve of 
\begin{equation}
\lambda_{\mbox{\scriptsize max}} = \alpha T^\beta,
\label{pow}\end{equation}
with $\beta \sim 0.38$. 
The data for the system of $\rho=1.2$ show similar dependences 
on temperature. 
Only a small deviation in $\lambda_{\mbox{\scriptsize max}}$ curve 
from the power function  of Eq.(\ref{pow}) (with $\beta=0.43$)
can be noticed over a wide range of 
$0.2 \, \raisebox{-.5ex}{$\stackrel{\textstyle <}{\sim}$} \,
 T \,\raisebox{-.5ex}{$\stackrel{\textstyle <}{\sim}$} \, 0.7$.

Such a smooth dependence of $E/N$ on temperature may imply that 
there is no phase transition at all or, if any, 
that it is of higher than first order. 
To investigate further
we considered the specific heat defined as 
\begin{equation}
C_V \equiv \frac{d(E/N)}{dT}, 
\end{equation}
and the first derivative of $\lambda_{\mbox{\scriptsize max}}$ 
with respect to temperature $d\lambda_{\mbox{\scriptsize max}}/dT$. 
From the obtained data for $E/N$ \it versus \rm $T$
and $\lambda_{\mbox{\scriptsize max}}$ \it versus \rm $T$,
the derivatives can be evaluated numerically, for example, 
by using the Lagrange's three-point interpolation formula.

In Fig. 3, the resulting derivatives are presented as a function 
of the temperature for the system of $\rho=1.0$. 
We have used three data points with  
$\Delta(E/N)=0.15$ throughout. 
The error bars in the figures were estimated 
by comparing the results of $\tau=1000$ and $\tau=2000$ as mentioned above.
In Fig. 3(a), for all the temperatures    
the specific heat has almost a constant value ($\sim 1.45$),
which is larger than that of the two-dimensional ideal gas 
($C_V/N=1$) but smaller than that of the ideal solid ($C_V/N=2$).
The temperature dependence of 
$d\lambda_{\mbox{\scriptsize max}}/dT$ is trivial;
it decreases monotonically as temperature increases. 
We conclude that in case of $\rho=1.0$ 
the system stays in a single phase, for all the temperatrures
in the range studied.

Fig. 4(a) and (b), respectively, are the specific heat and 
$d\lambda_{\mbox{\scriptsize max}}/dT$ for the system of $\rho=1.2$. 
The specific heat exhibits a $\lambda$-shaped peak 
around $T \sim .56$. 
The first derivative, $d\lambda_{\mbox{\scriptsize max}}/dT$,
shows also a peak at the {\em same} temperature 
on top of the monotonically decreasing background curve. 
Although the finite-size effects and the 
large $\Delta T$ used in the numerical differentiation 
would have made the shapes of the peaks less sharp, 
those peaks are sufficiently well-defined to support our conclusion that 
the system undergoes a solid-fluid phase transition 
of the second-order at the corresponding temperature.

%%%%%%%%%%%%%%%%%%%%%%%%%%%%%%%%%%%%%%%%%%%%%%%%%%%%%%%%%%%%%%%%%%%%%%
%%%%%%%%%%%%%%%%%%%%% Three-Dimensional System %%%%%%%%%%%%%%%%%%%%%%%
%%%%%%%%%%%%%%%%%%%%%%%%%%%%%%%%%%%%%%%%%%%%%%%%%%%%%%%%%%%%%%%%%%%%%%

\section{Three-Dimensional System}
The three-dimensional system studied here is $N$=32 interacting 
particles moving in a periodic cubic box. 
The periodic box contains $8(=2\times 2 \times 2)$ primitive cells 
of the face-centered-cubic(fcc) lattice when
the particles are in the configuration of the minimum 
potential energy.
Since the volume of the cubic primitive cell containing four particles is 
$(\sqrt2 d)^3$ with the lattice constant $d$, the
particle number density $\rho$ is determined simply  
as $\rho = \sqrt2/d^3$.

In Fig. 5, the same quantities studied in Sec. IV
(that is,  $E/N$, $\lambda_{\mbox{\scriptsize max}}$ 
and their derivatives with respect to the temperature) are 
presented as a function of the temperature for the 
three-dimensional system with particle density $\rho=1.0$.
The temperature is now defined as 
$T \equiv \frac32 (\langle K\rangle/N)$.  
The averages are also taken for $\tau=2000$ after discarding data from the 
initial time interval of length $t=100$.   
At this density, the system energy per particle $E/N$ shows a trivial 
linear dependence on the temperature (Fig. 5a), which leads to
almost constant specific heat (Fig. 5c).
The  temperature dependence of $\lambda_{\mbox{\scriptsize max}}$ 
can be fitted into a power function of Eq.(\ref{pow}) 
with $\beta=0.37$. 
These facts suggest that the system of $\rho=1.2$ 
remains in a single phase, {\it i.e.}, the fluid phase. 

Fig. 6 are show $E/N$ \it versus \rm $T$ and 
$\lambda_{\mbox{\scriptsize max}}$ \it versus \rm $T$ for the systems of the 
particle density $\rho=1.3$(a,c) and $\rho=1.4$(b,d).  
In case of $\rho=1.3$, 
discrete jumps can be seen both in the $E/N$ \it versus \rm $T$ 
graph and in the $\lambda_{\mbox{\scriptsize max}}$ \it versus \rm $T$ 
graph at the same temperature $T \sim 0.14$.  
Note that this already happens below $\rho_c=\sqrt2$, 
at which density the particles begin to contact each other 
in the minimum energy fcc configuration. (The subcript `c' 
refers to `contact'.)
A similar discrete jump can be seen also in the system with $\rho=1.4$ 
at higher temperature.  
Furthermore, over a wide range of temperature  
($0.26 \, \raisebox{-.5ex}{$\stackrel{\textstyle <}{\sim}$} \,
 T \,\raisebox{-.5ex}{$\stackrel{\textstyle <}{\sim}$} \, 0.32$. 
the averages taken over $\tau=2000$ show large deviations 
from those over $\tau=1000$.

Such poor convergence of the averages indicates that 
the system is unstable near the phase transition. 
In Fig. 7, we present the average values 
$T(\equiv \frac23 \langle K\rangle/N)$ 
and $\lambda_{\mbox{\scriptsize max}}$ as a function of $\tau$ 
over which the averages are taken.  
The dashed-dotted curves 
are averages for the system of $E/N=0.5$ 
which show a good convergence to stable constant values. 
Especially, the convergence of the averaged kinetic energy 
is remarkably rapid compared to the Lyapunov exponent. 
Such a standard behavior holds for all the averages that have been 
discussed up to now. Furthermore, the resulting averages 
do not depend on the initial conditions.
On the other hand, the solid curves are the averages 
for the system of $E/N=0.7$. 
The averages seem to reach somewhat stable values up to  
$\tau\sim 600$, at which point the averages suddenly depart from 
those trends.  
They also show a strong dependence on the initial conditions 
of the simulations. The dashed curves are the corresponding 
quantities for the system of the same energy $E/N=0.7$ but 
started from different set of initial velocities. 
At first, the averages tends to converge to {\em different} values
from those of the solid curves 
till the sudden changes occur again.
The final values of the solid and dashed curves at $\tau=2000$ 
seem to converge to the same value but they still show a large difference.

In order to understand such a peculiar behavior in taking the 
averages, we introduce `local' averages in the vicinity of $t$ 
defined as 
\begin{equation}
\langle A\rangle(t) \equiv \frac{1}{\Delta \tau} 
\int^{t+\frac12\Delta\tau}_{t-\frac12\Delta\tau} dt^\prime A(t^\prime).
\end{equation}
In Fig. 8, shown are the local temperature and Lyapunov exponents 
that lead to the accumulated averages shown in Fig.7 (solid and dash-dotted 
curves).
The local averages for the system of $E/N=0.5$ 
oscillate about a single value, 
resulting in a stable cumulative average.
In case of $E/N=0.7$, there appear instead to be two centers of
oscillation for  
the local averages, indicating that the system has two 
`quasi-stable' phases.
The system cannot stay in either phase but instead changes between
the phases 
from time to time. 
Note that the transfer occurs abruptly because the system is relatively small.

How long the system stays in one phase is senstive to initial conditions;
that is, a tiny difference in the initial conditions 
or any small fluctuations coming from roundoff errors in the 
numerical simulations leads to quite different results.
This interpretation explains not only the abrupt changes
in the accumulated averages 
in Fig. 7 but also the strong dependence of them on the 
initial conditions. 
Thus, in order to obtain averages with a few-digit accuracy 
we have to simulate the motion for a large $\tau$  
in the phase-transition region. 
Shown in Fig. 9 are the distributions of local averages 
obtained with $\Delta \tau=5$ for 20000 time units (ten times 
longer than what we have used in evaluating the averages discussed 
so far). 
In case of $E/N=0.5$ the distribution of the local temperature 
and the local Lyapunov exponents (black bars) can be fitted into 
single Gaussian curves (solid lines). 
On the other hand, those for the system in the phase 
transition region($E/N=0.7$, white bars) 
split into two Gaussian distributions. 
The dashed ones are the separate Gaussian curves 
and the solid one is their sum, which fits the whole distributions.
Those two Gaussian curves correspond to the 
quasistable fluid and solid phases of the system, respectively.

By noticing that the two Gaussian curves for the distribution 
of the local temperature is clearly divided into two, 
we can evaluate the averages of the quasi-stable phases separately.
To do this, we sum up the local temperature 
and local Lyapunov exponents for the time period when  
$T_{\mbox{\scriptsize local}} < T_{\mbox{\scriptsize sep.}}$
or $T_{\mbox{\scriptsize local}} > T_{\mbox{\scriptsize sep.}}$
with a properly chosen $T_{\mbox{\scriptsize sep.}}$.
(See Fig.9 for the definition of $T_{\mbox{\scriptsize sep.}}$.)
The genuine overall average of the system lies in between those two values. 
Shown in Fig.10 are those averages of the system obtained through a
similar analysis with the local averages for 
$\tau=20000$; that is, the overall averages (solid circles) 
and the averages of the quasistable phases (open circles). 
The error bars for the solid circles 
indicate the standard deviations of the local averages. 
The quasistable phases smoothly match 
with the averages of the system in a stable single phase. 
Thus, those quasi-stable phases can be interpreted as the 
``super-cooled" fluid phase and ``super-heated" solid phase 
in the literature.
The $E/N$ \it versus \rm $T$ data for the quasi-stable phases 
and the stable single phase can be fit into a straight line, 
$E/N=2.87 T - 0.09$ for the solid phase and $E/N=2.90 T + 0.08$ for 
the fluid phase around the phase transition region. 
It enables us to evaluate the latent heat (per particle) associated 
with the phase transition as  
$\Delta \ell\sim 0.17(=0.08+0.09)$. 
Mixing phenomena in the hard sphere system have been investigated
by using the maximum Lyapunov exponents in Ref.\cite{DP97}.

Although the data we obtained still show large standard deviations, 
they are sufficiently precise for us to see what happens in our 3-dimensional 
system when it undergoes a phase transition. That is, 
our three-dimensional system undergoes a first-order phase transition. 
The system energy and the largest Lyapunov exponent 
discontinuously jump at the temperature of the 
phase transition.

%%%%%%%%%%%%%%%%%%%%%%%%%%%%%%%%%%%%%%%%%%%%%%%%%%%%%%%%%
%%%%%%%%%%%%%%%%%%%%% Conclusions %%%%%%%%%%%%%%%%%%%%%%%
%%%%%%%%%%%%%%%%%%%%%%%%%%%%%%%%%%%%%%%%%%%%%%%%%%%%%%%%%

\section{Conclusions}
In this work, we studied changes in the chaos of a 
many-body system as that system underwent a phase transition. 
As models we considered $N$-particles moving in 
two-dimensional ($N=30$) or three-dimensional ($N=32$) periodic 
boxes. The particles interacted with each other through a short-ranged
repulsive 
potential with a soft-core.  
We computed the largest Lyapunov exponents for the motions of the particles
as a function of temperature for a few different densities.  
Relaxation phenomena in the phase transition region 
were analyzed by using the local time averages.
In a conclusion, the Lyapunov exponent can be a 
good physical quantity for investigating the phase transition. 
It may tell us {\em when the phase transition occurs} and 
{\em what kind of phase transition is involved.}

Here we have considered only a simple soft-core potential 
only with the repulsive interactions. 
Furthermore, we have carried out the simulation with a fixed system 
energy.  
To see whether or not the difference in transition order
is more general,
we will have to check it carefully 
for the other systems with more realistic interactions 
and/or for the thermostatted systems with a fixed temperature.
Work in this direction is under progress and will be reported 
elsewhere\cite{WiP}.

%%%%%%%%%%%%%%%%%%%%%%%%%%%%%%%%%%%%%%%%%%%%%%%%%%%%%%%%%%%%%%%%
%%%%%%%%%%%%%%%%%%%%% Acknowledgements %%%%%%%%%%%%%%%%%%%%%%%%%
%%%%%%%%%%%%%%%%%%%%%%%%%%%%%%%%%%%%%%%%%%%%%%%%%%%%%%%%%%%%%%%%

\acknowledgements
We are grateful for the discussions with Professor Wm. G. Hoover 
during his recent visit to Korea and for his reading the manuscripit 
with a lot of helpful comments. 

%%%%%%%%%%%%%%%%%%%%%%%%%%%%%%%%%%%%%%%%%%%%%%%%%%%%%%%%%%%%%%%%
%%%%%%%%%%%%%%%%%%%%%%%%% References %%%%%%%%%%%%%%%%%%%%%%%%%%%
%%%%%%%%%%%%%%%%%%%%%%%%%%%%%%%%%%%%%%%%%%%%%%%%%%%%%%%%%%%%%%%%

\newpage

\begin{figure}
\centerline{\epsfig{file=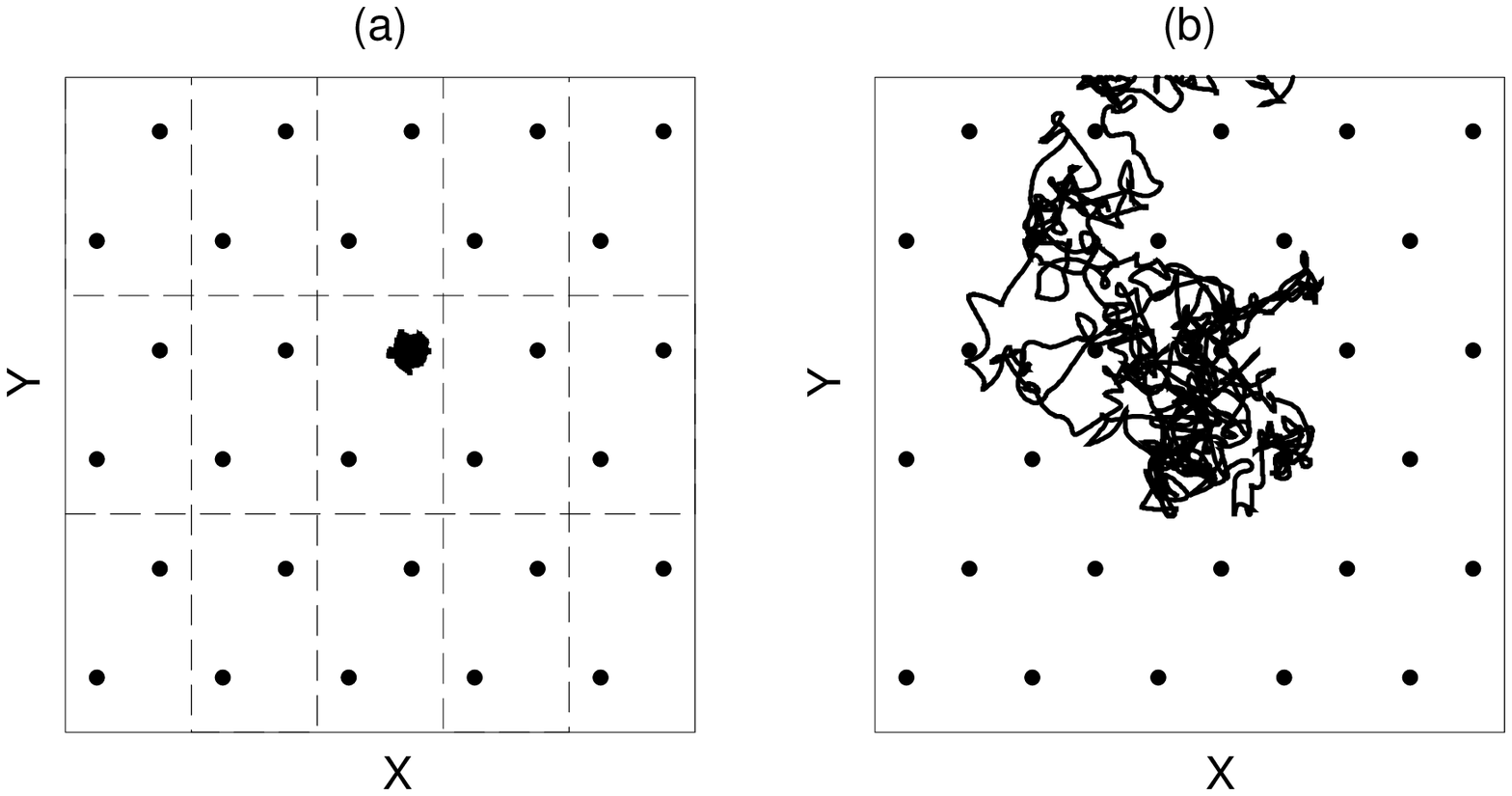, width=16cm}}
\caption{
Two characteristic single-particle motions of the system 
(a) in the solid phase and (b) in the fluid phase.}
\end{figure}

\begin{figure}
\centerline{\epsfig{file=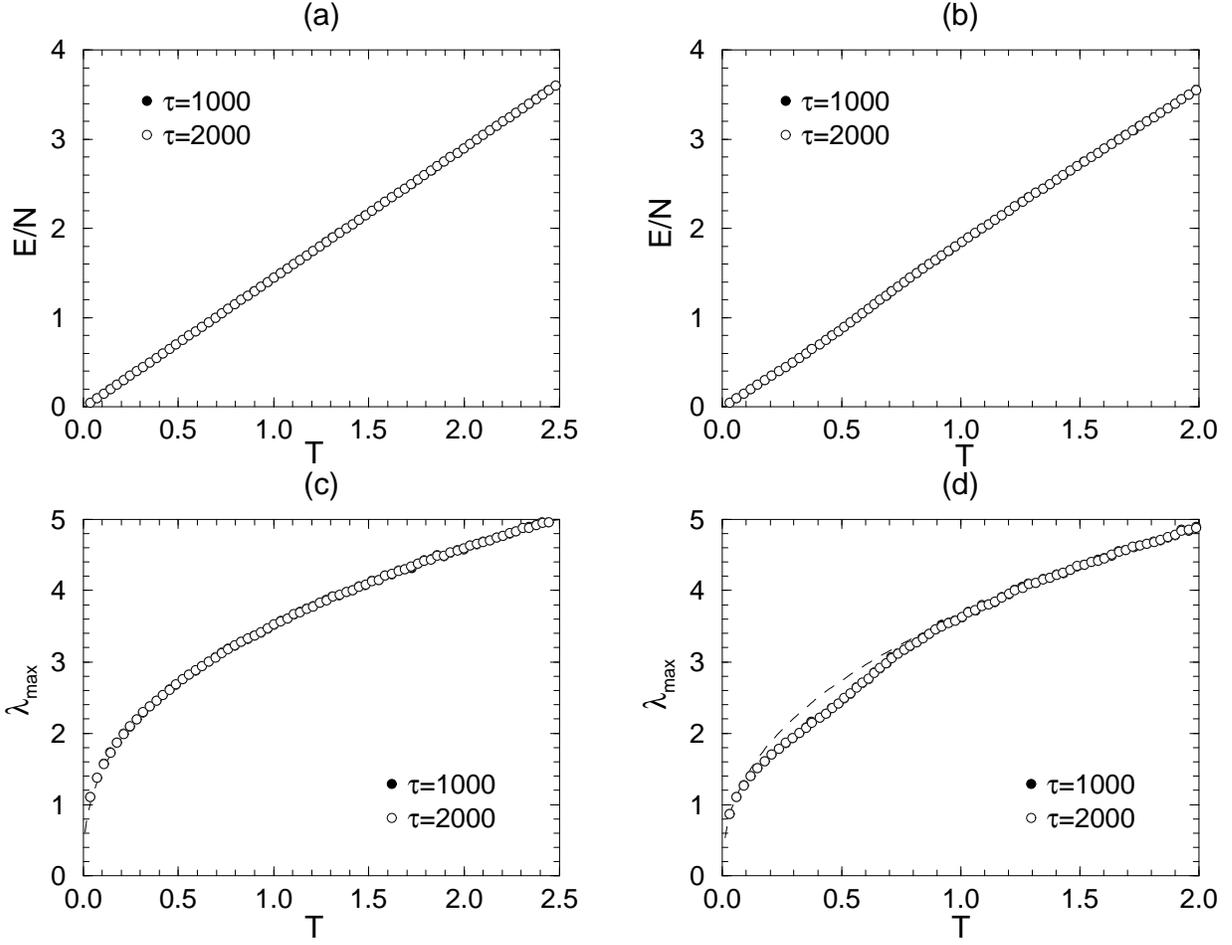, width=17cm} }
\caption{$E/N$ \it versus \rm temperature($T$) and 
$\lambda_{\mbox{\scriptsize max}}$ \it versus \rm T for the system with 
$\rho=1.0$ (a,b) and $\rho=1.2$ (c,d).}
\end{figure}

\newpage
\begin{figure}
\centerline{\epsfig{file=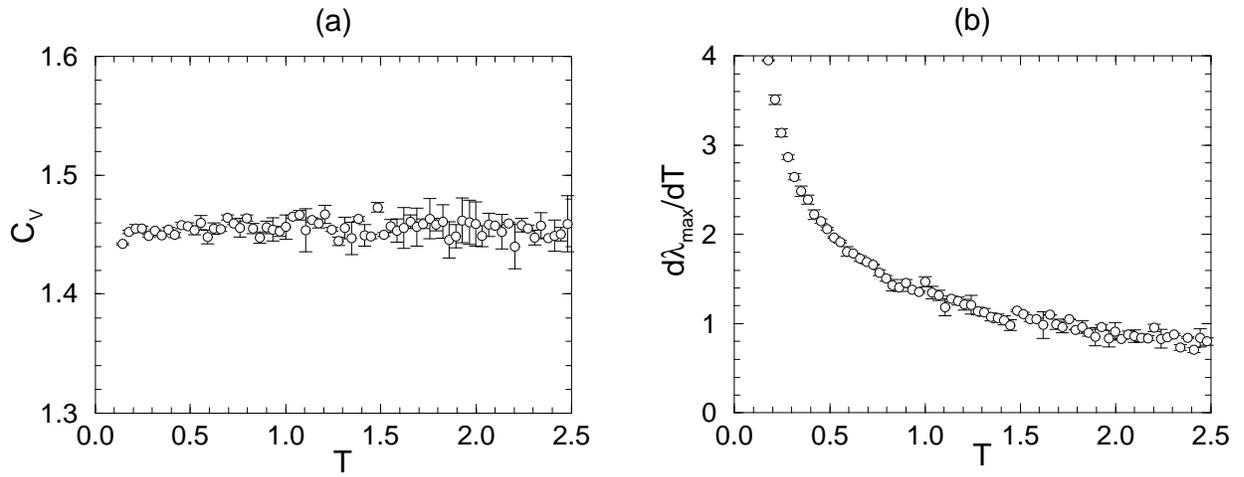, width=17cm} }
\caption{The specific $C_V$ and 
$d\lambda_{\mbox{\scriptsize max}}/dT$ 
\it versus \rm temperature for the system of $\rho=1.0$.  }
\end{figure}

\vskip 2cm
\begin{figure}
\centerline{\epsfig{file=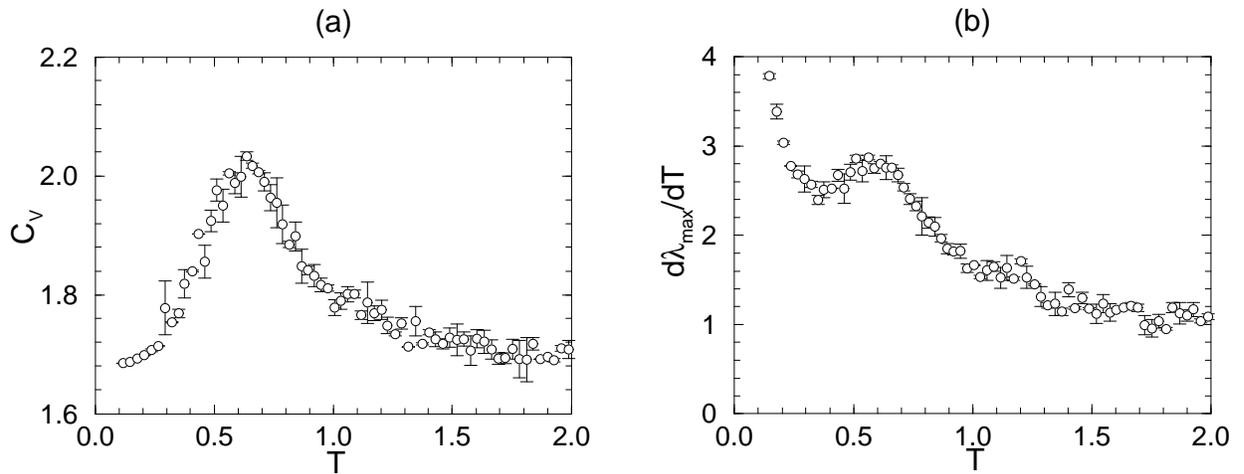, width=17cm} }
\caption{Same quantities as Fig. 3 for the system of $\rho=1.2$.}
\end{figure}

\begin{figure}
\centerline{\epsfig{file=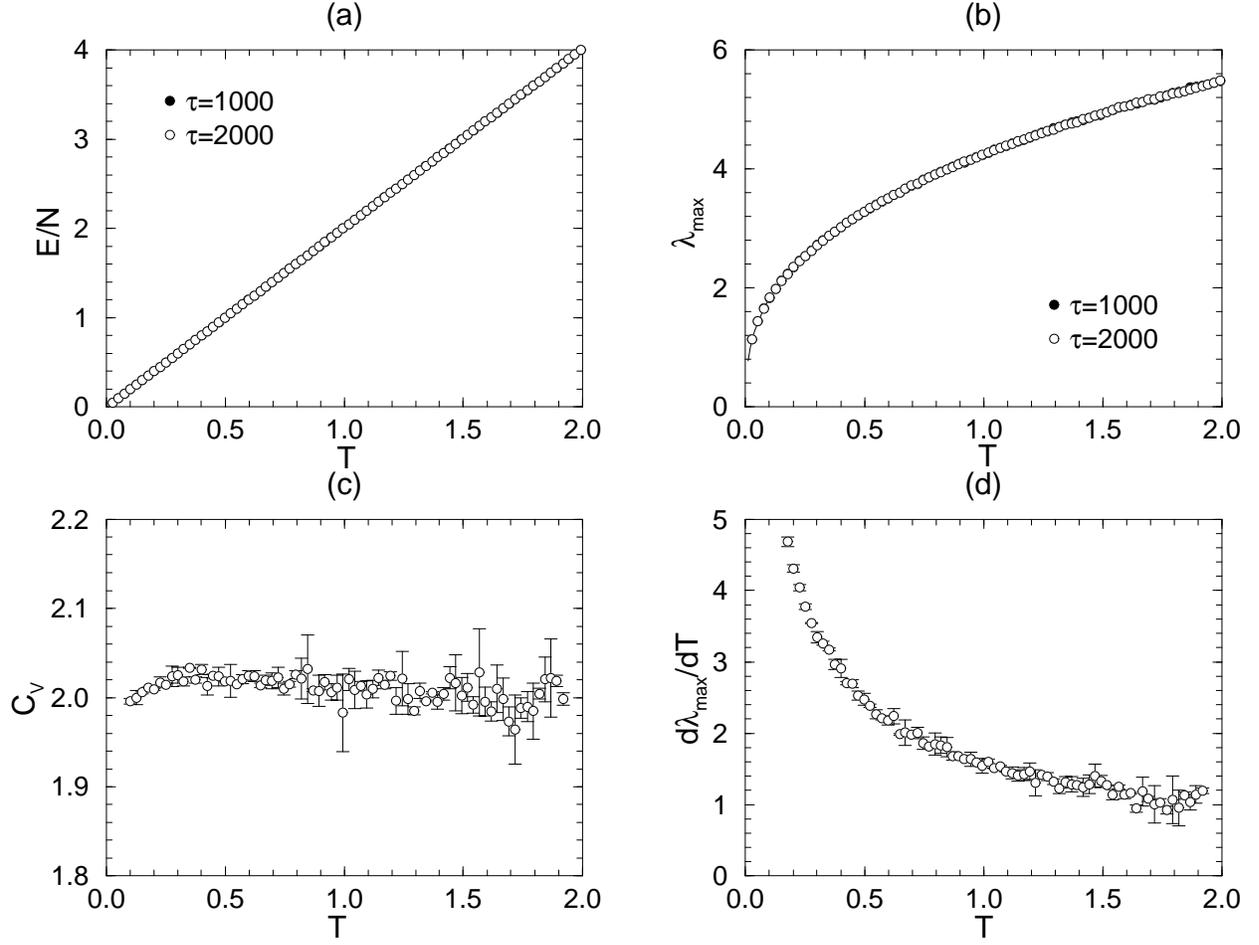, width=17cm} }
\caption{(a) $E/N$ \it versus \rm temperature($T$), 
(b) $\lambda_{\mbox{\scriptsize max}}$ \it versus \rm $T$, 
(c) $C_V$ \it versus \rm $T$, 
and (d) $d\lambda_{\mbox{\scriptsize max}}/dT$ \it versus \rm T 
for a three-dimensional system with the particle density  
$\rho=1.0$.}
\end{figure}

\begin{figure}
\centerline{\epsfig{file=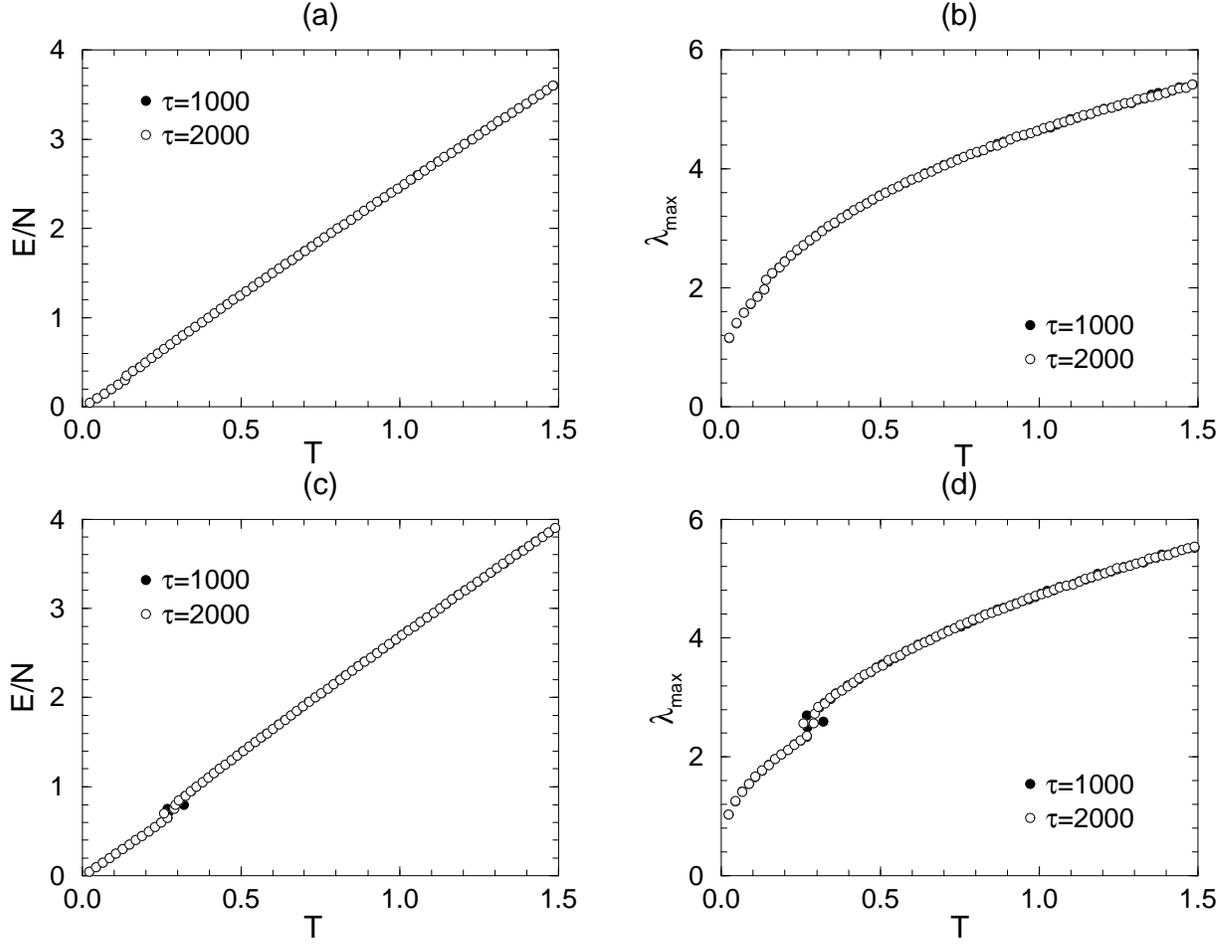, width=17cm} }
\caption{(a) $E/N$ \it versus \rm temperature($T$), 
(b) $\lambda_{\mbox{\scriptsize max}}$ \it versus \rm $T$ for the system of 
particle density $\rho=1.3$, and  
(c) $E/N$ \it versus \rm $T$, 
and (d) $\lambda_{\mbox{\scriptsize max}}$ \it versus \rm T 
for the system with $\rho=1.4$. }
\end{figure}

\newpage
\begin{figure}
\centerline{\epsfig{file=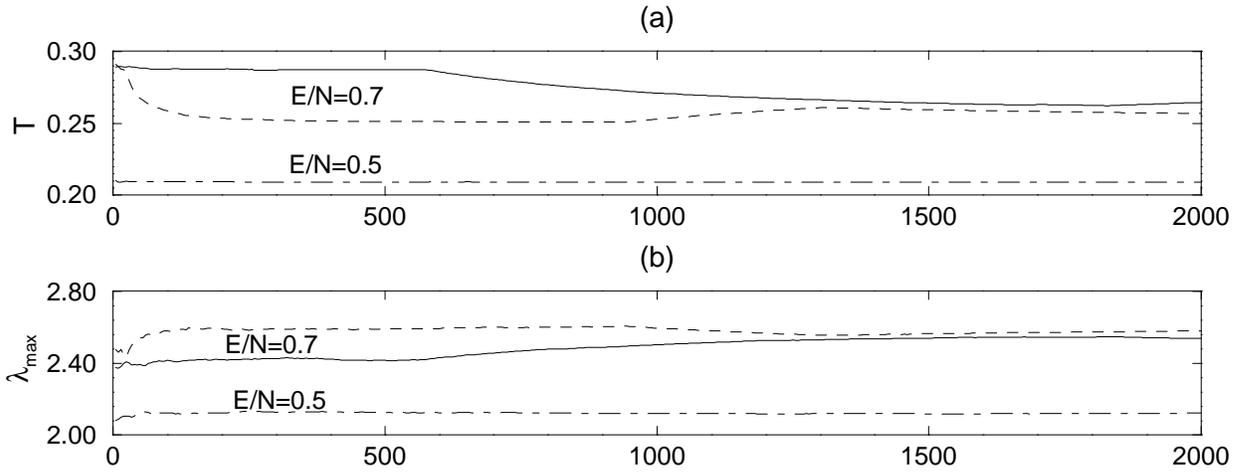, width=16cm}}
\vskip 3mm 
\caption{Poor convergence of averages near the phase transition :
(a) $T(\equiv \frac32\langle K\rangle/N)$ and 
(b) $\lambda_{\mbox{\scriptsize max}}$ as a function of $\tau$ for the 
system of $\rho=1.4$ with $E/N=0.7$ (solid lines). 
The dash-dotted lines are the corresponding averages for the system 
of the same particle density but with $E/N=0.5$, 
which converge rapidly to stable values.}
\end{figure}

\vskip 2cm
\begin{figure}
\centerline{\epsfig{file=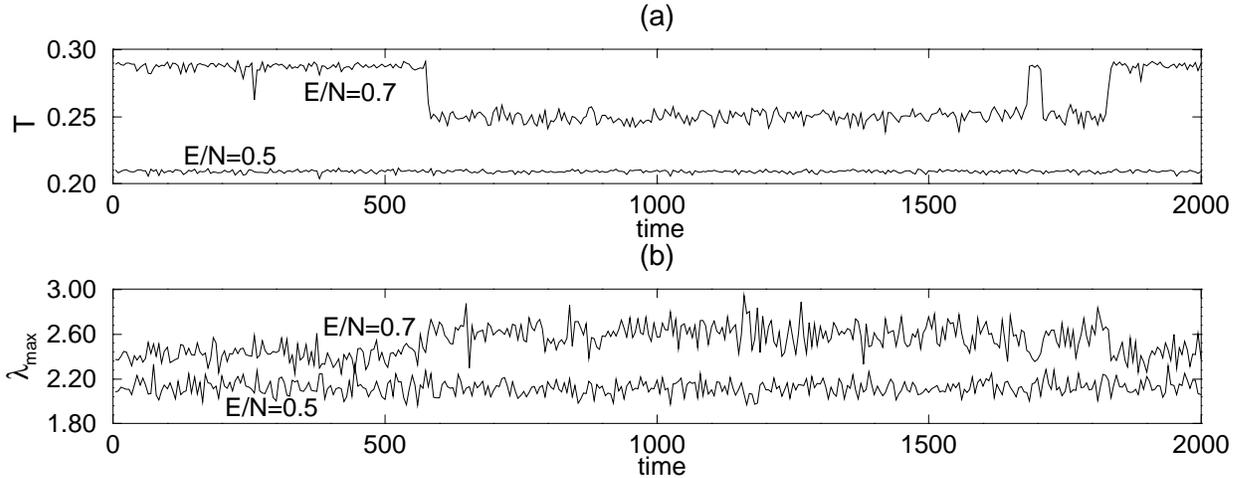, width=16cm}}
\vskip 3mm 
\caption{Local averages with $\Delta\tau=5$ for the system 
of density $\rho=1.4$ and energy $E/N=0.5$ and 0.7 :
(a) $T(\equiv \frac32\langle K\rangle/N)$ and 
(b) $\lambda_{\mbox{\scriptsize max}}$.}
\end{figure}

\newpage
\begin{figure}
\centerline{\epsfig{file=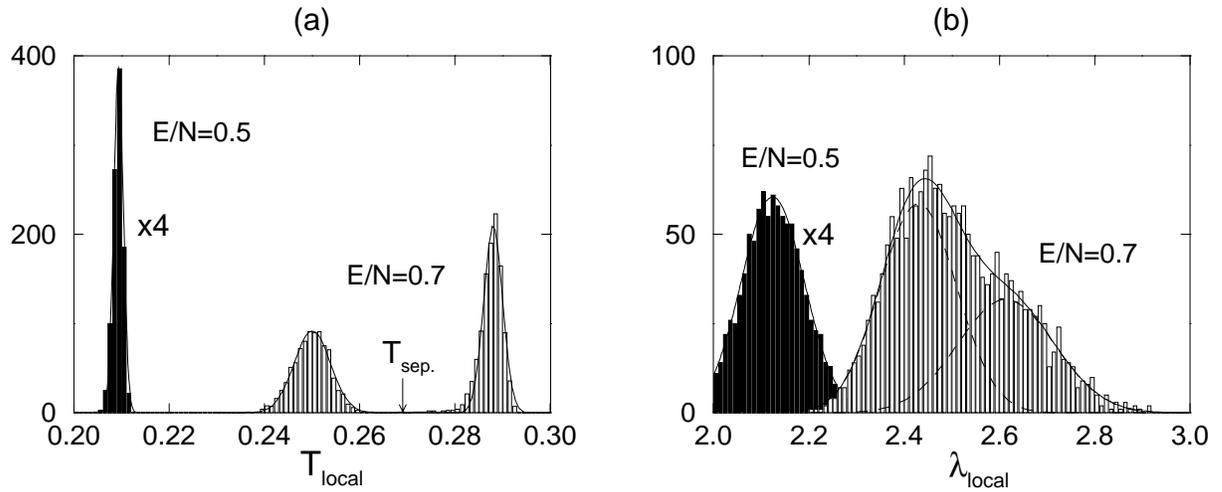, width=17cm}}
\vskip 3mm 
\caption{Distribution of local averages with $\Delta\tau=5$ for $\tau=20000$
for the system 
of density $\rho=1.4$ and energy $E/N=0.5$ and 0.7 :
(a) $T(\equiv \frac32\langle K\rangle/N)$ and 
(b) $\lambda_{\mbox{\scriptsize max}}$.}
\end{figure}

\vskip 2cm
\begin{figure}
\centerline{\epsfig{file=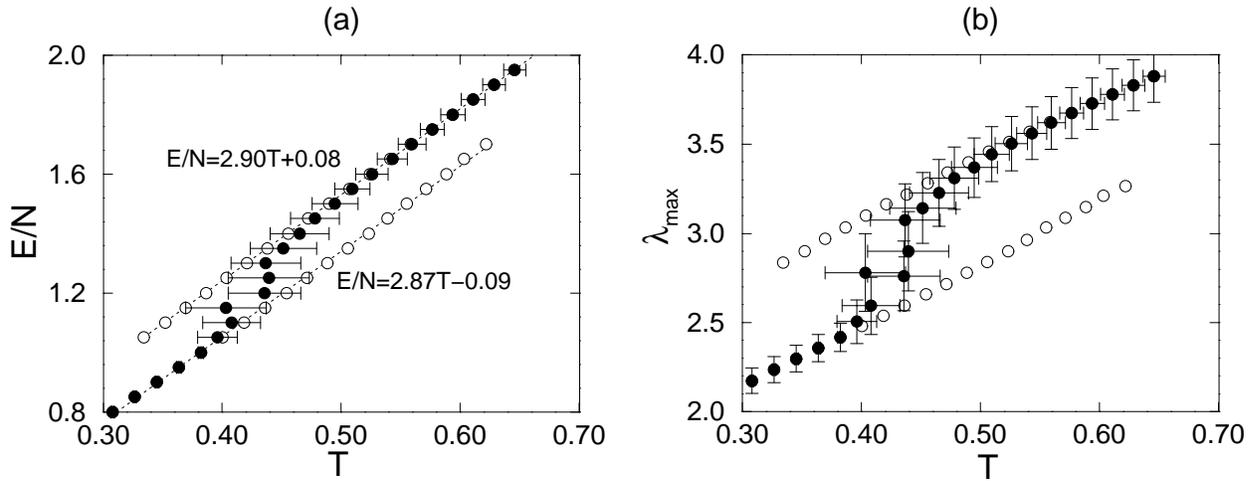, width=17cm}}
\vskip 3mm 
\caption{(a) $E/N$ \it versus \rm $T$ and 
(b) $\lambda_{\mbox{\scriptsize max}}$ \it versus \rm $T$ near the phase 
transition. The solid circles are the overall accumulated 
averages of the system for $\tau=20000$ 
and the open circles are averages of the quasi-stable 
phases obtained in the way described in the text.}
\end{figure}

\end{document}